\documentclass[aps,onecolumn,preprint,groupedaddress,superscriptaddress,showpacs]{revtex4-1}

\usepackage{graphicx}
\usepackage{color}
\usepackage{amsmath}
\usepackage{amsfonts}
\usepackage{amssymb}
\usepackage{dcolumn}
\usepackage{hyperref}
\usepackage{bbold}
\usepackage{xr}
\begin{document}
  \title{Temporal dynamics of online petitions}

 \author{L. B\"{o}ttcher}
 \email{lucasb@ethz.ch}
 \affiliation{ETH Zurich, Wolfgang-Pauli-Strasse 27, CH-8093 Zurich,
Switzerland}  
\author{O. Woolley-Meza}
 \affiliation{ETH Zurich, Clausiusstrasse 50, CH-8092 Zurich,
Switzerland}
\author{D. Brockmann}
\affiliation{Humboldt Universit\"{a}t zu Berlin, Philippestrasse 13, Building 4, 10115 Berlin, Germany}
\affiliation{Robert Koch-Institute, Nordufer 20, 13353 Berlin, Germany}
\date{\today}
\begin{abstract}
Online petitions are an important avenue for direct political action, yet the dynamics that determine when a petition will be successful are not well understood. Here we analyze the temporal characteristics of online-petition signing behavior in order to identify systematic differences between popular petitions, which receive a high volume of signatures, and unpopular ones. We find that, in line with other temporal characterizations of human activity, the signing process is typically non-Poissonian and non-homogeneous in time. However, this process exhibits anomalously high memory for human activity, possibly indicating that synchronized external influence or contagion play and important role. More interestingly, we find clear differences in the characteristics of the inter-event time distributions depending on the total number of signatures that petitions receive, independently of the total duration of the petitions. Specifically, popular petitions that attract a large volume of signatures exhibit more variance in the distribution of inter-event times than unpopular petitions with only a few signatures, which could be considered an indication that the former are more bursty. However, petitions with large signature volume are less bursty according to measures that consider the time ordering of inter-event times. Our results, therefore, emphasize the importance of
accounting for time ordering to characterize human activity.
\end{abstract}
\maketitle
\section{Introduction}
In the last decade online activism \cite{cyberactivism}, in particular online petition platforms such as \emph{openPetition}, \emph{Avaaz}, \emph{change.org}, \emph{Campact}, \emph{MoveOn.org} attracted the attention of many citizens since these platforms allow a more active participation in politics and also enable them to form communities for getting involved in local or even international political decisions. Despite the great interest in online activism, relatively little is known about benefit, success and the dynamics of online petitions \cite{yasseri13,hale13}. Here we analyze the temporal evolution of petitions belonging to \emph{openPetition}, an online platform that aims to support citizens to make their requests public, coordinate their actions and open a dialog with elected representatives. Geographically, the platform addresses online petitions from Germany, Austria, Switzerland and member countries of the European Union. For every petition \emph{openPetition} calculates a quorum defining the minimum number of signatures needed to make the platform sending a request for an official statement from the responsible representatives. However, regardless whether the petition reaches quorum, it can be always handed over. Since the launch of the \emph{openPetition} homepage in 2010 more than 3 million users signed over 14 million times.

Here we analyze the temporal characteristics of signing on the \emph{openPetition} platform. Previous studies investigating the temporal features of human activity have reported a power-law distribution of inter-event times \cite{barabasi05,didomenico13}. These findings are in sharp contrast to the general assumption of exponentially distributed inter-event times, i.e.~a Poisson process description of human activity \cite{daley88,anderson03}. However, different mechanisms have been suggested to explain deviations from a Poisson description \cite{barabasi05,Malmgren08,jo12}. We show that online petition signing time-series exhibit different characteristic inter-event time distributions according to their popularity, i.e.~the total number of signatures a petition receives.
We further analyze this effect by studying regularity and burstiness of the signing ``spike trains'', i.e.~the time series of discrete signing events.
Burstiness is, broadly speaking, intermittent activity. However, the specific definition varies. In the study of human activity it was initially defined as short intervals of high activity followed by log intervals of inactivity~\cite{goh08}. However, subsequently measures of burstiness from the neuroscience literature, specifically the local variation, have proved useful in characterizing human activity \cite{shinomoto03,omi11,sanli15,sanli2015,lambiotte16}. Originally defined to describe neuronal bursts, in this context a bursty time-series exhibits short inter-event times typically followed by long ones.
% As one example, the analysis of Twitter hashtag activity based on the local variation, a measure considering the time ordering of inter-event times, has been proven useful to reveal characteristic temporal features of hashtags with high or low activity \cite{sanli15,sanli2015}. 
We apply different measures of burstiness to characterize online petition time series by studying the local variation of the signing spike trains. These measures are introduced in the \emph{Materials and Methods} section. In the subsequent \emph{Results} section we discuss the characterization of the petitions' temporal features. Most importantly, our characterization allows for a clear differentiation between popular petitions that accrue a large volume of signatures and less popular ones that fail to do so. As described in the concluding \emph{Discussion} section, this could point to a fundamental difference in the signing dynamics at work in each case. Furthermore, our results emphasize the importance of burstiness measures, such as the local variation, that take the time ordering of inter-event times into account.
\section{Materials and methods}

The data set we analyze contains detailed information on all petitions of the online platform \emph{openPetition}. Besides petition name, content, category and current status, the data base also includes signers' and initiators' geolocations and the corresponding signature timestamps. In total, there are 16282 petitions with 10948145 signatures. 10570 petitions have at least one signature. The most recent numbers are available online: \url{https://www.openpetition.de/}. In this study, we focus on the analysis of the petitions' time series which are based on the corresponding time stamps. To investigate their regularity and burstiness we use different measures which we describe below.

The local variation $L_V$ has been applied to study the spiking characteristics of non-stationary processes, in particular for investigating neural spike trains \cite{shinomoto03,omi11}, and is defined by:
\begin{equation}
\label{eq:lv}
L_V= \frac{3}{N-2}\sum_{i=2}^{N-1}\left(\frac{\tau_{i+1}-\tau_i}{\tau_{i+1}+\tau_i}\right)^2,
\end{equation}
where $\tau_i$ corresponds to the $i$-th inter-event time.
Recently, the local variation has been used to analyze temporal features of Twitter hashtags \cite{sanli15,sanli2015}. $L_V$ takes values within the interval $[0,3]$ and approaches unity for Poisson sequences. This can be seen by assuming a gamma distribution $p_{\kappa}\left( \tau \right)=\left( \kappa \xi \right)^\kappa \tau^{\kappa -1} \exp\left(-\kappa \xi  \tau\right)/\Gamma\left( \kappa \right)$, where $\kappa$ denotes the shape parameter controlling the burstiness (irregularity) and $\xi$ defines the firing rate determining the speed of spike train dynamics. Calculating the average of $L_V$ with respect to the given distribution $p_{\kappa}(\tau)$ yields \cite{shinomoto03}:
\begin{align}
&\langle L_V \rangle = \int_0^{\infty}d \tau_1 \int_0^{\infty}d \tau_2 3 \frac{\left(\tau_1-\tau_2 \right)^2}{\left(\tau_1 + \tau_2 \right)^2} p_{\kappa}(\tau_1)p_{\kappa}(\tau_2)=\frac{3}{2 \kappa + 1}.
\end{align}
For the gamma distribution $p_\kappa(\tau)$, mean and standard deviation are given by $m_\tau=1/\xi$ and $\sigma_\tau=m_\tau/\sqrt{\kappa}$ respectively. For $\kappa = 1$ we obtain a Poisson process and indeed $L_V=1$ \cite{omi11}. The standard deviation $\sigma_\tau$ of the signal increases as $\kappa$ decreases---the signal is said to be more bursty. Thus $L_V>1$ indicates that the signal is more bursty than a signal generated by a homogeneous Poisson process and $L_V<1$ one that is less. Deviations from the Poissonian signal ($L_V=1$) either occur because of a non-exponential inter-event time distribution or are due to correlations in the signal.

Besides the local variation $L_V$, we also consider two other measures in this study, namely the burstiness coefficient $B$ and the memory coefficient $M$ \cite{goh08}.
%Prior work claims allow a clear distinction of time series generated human activity from the ones originating from natural phenomena and texts \cite{goh08}.
The burstiness coefficient $B$ is defined as:
\begin{equation}
\label{eq:burstiness}
B=\frac{\sigma_{\tau}/m_{\tau}-1}{\sigma_{\tau}/m_{\tau}+1}=\frac{\sigma_{\tau}-m_{\tau}}{\sigma_{\tau}+m_{\tau}},
\end{equation}
where $m_{\tau}$ is the mean of the inter-event time distribution and $\sigma_{\tau}$ the corresponding standard deviation. For real-world finite time series with existing mean and standard deviation the values of $B$ are within the interval $(-1,1)$ \cite{goh08}. An exponential inter-event time distribution with $\kappa=1$ yields $B=0$. For the most bursty signal for which the variance approaches infinity, we find $B=1$. Completely regular signals are described by $B=-1$. In contrast to $L_V$, the measure $B$ does not take into account the temporal order of inter-event times.

We characterize the correlation properties of a time series using the memory coefficient $M$, which is simply the correlation coefficient of consecutive inter-event times $(\tau_i,\tau_{i+1})$:
\begin{equation}
\label{eq:memory}
M=\frac{1}{n_{\tau}-1}\sum_{i=1}^{n_{\tau}-1} \frac{\left(\tau_i -m_1 \right)\left(\tau_{i+1} -m_2 \right)}{\sigma_1 \sigma_2},
\end{equation}
where $n_{\tau}$ is the total number of inter-event times. The mean and the standard deviation of all inter-event times excluding the last one (the first one) are denoted by $m_1$ and $\sigma_1$ ($m_2$ and $\sigma_2$) respectively. The memory coefficient takes values in the interval $(-1,1)$ and is positive for signals where short (long) inter-event times have a tendency to be followed by another short (long) one, and it is negative in the opposite case.

\pagebreak
\section*{Results}
\label{sec:results}
\subsection{Time evolution of petitions' numbers of signatures and their distribution}
\begin{figure}
\centering
\includegraphics[width=\textwidth]{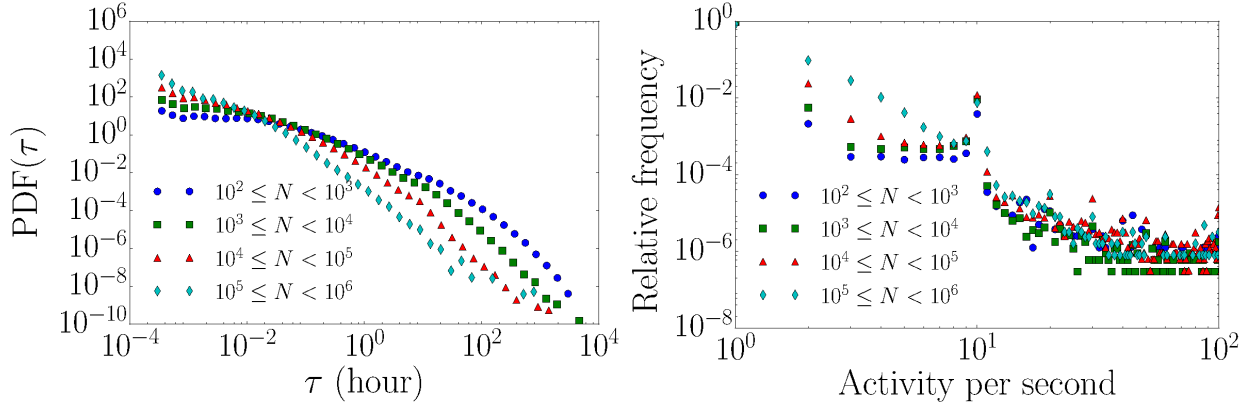}
  \caption{\textbf{Inter-event time distribution and burstiness.} All petitions are divided into four different classes based on the number of signatures $N$. (left) The corresponding probability density function (PDF) of the inter-event time intervals (hour). (right) The relative frequency of signing activity per second. The vast majority of signing activity corresponds to one signing event per time stamp.}
 \label{fig:tau_pdf}
\end{figure}
\begin{figure}
\centering
\includegraphics[width=\textwidth]{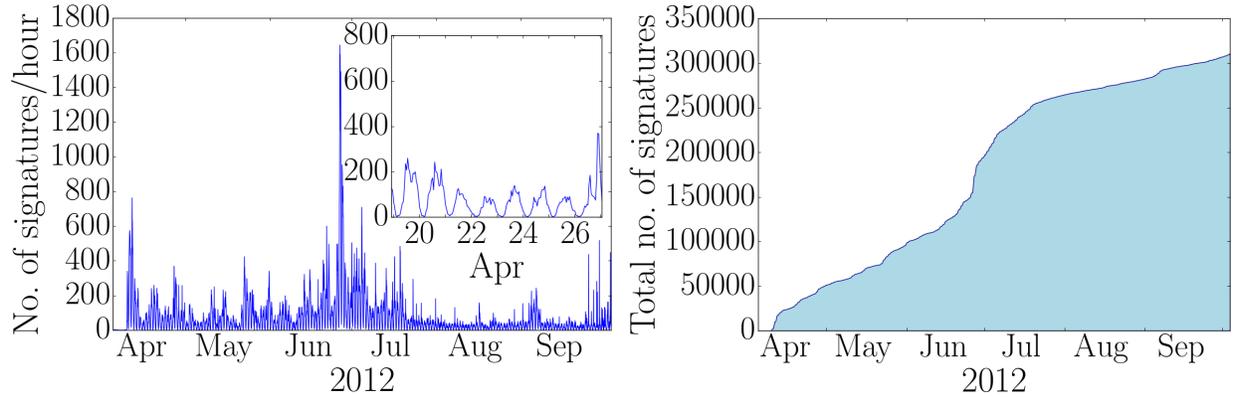}
  \caption{\textbf{Signing time series and time evolution of total number of signatures.} (left) Time series of the largest petition's signing activity per hour. The inset shows the superimposed circadian pattern. (right) The corresponding total number of signatures as a function of time.}
 \label{fig:spikes}
\end{figure}

In order to understand the temporal features of successful petitions with a large number of signatures, we divide all petitions into four different classes based on the number of signatures $N$, cf.~Fig.~\ref{fig:tau_pdf}. In subsequent paragraphs we refer to them as signature number classes. Class 1 contains petitions with the smallest numbers of signatures and class 4 the ones with the largest numbers respectively. The numbers of petitions in each class are: 2182, 1213, 151 and 8 (ascending class index). The probability density function of the inter-event times is shown in Fig.~\ref{fig:tau_pdf} (left). As expected, inter-event times are more broadly distributed than the exponential distribution expected from a Poisson process. In Fig.~\ref{fig:tau_pdf} (right), we show the relative frequency of signing activity per second. Most of the signing activity corresponds to one signing event per time stamp.
\begin{figure}
\centering
\includegraphics[width=\textwidth]{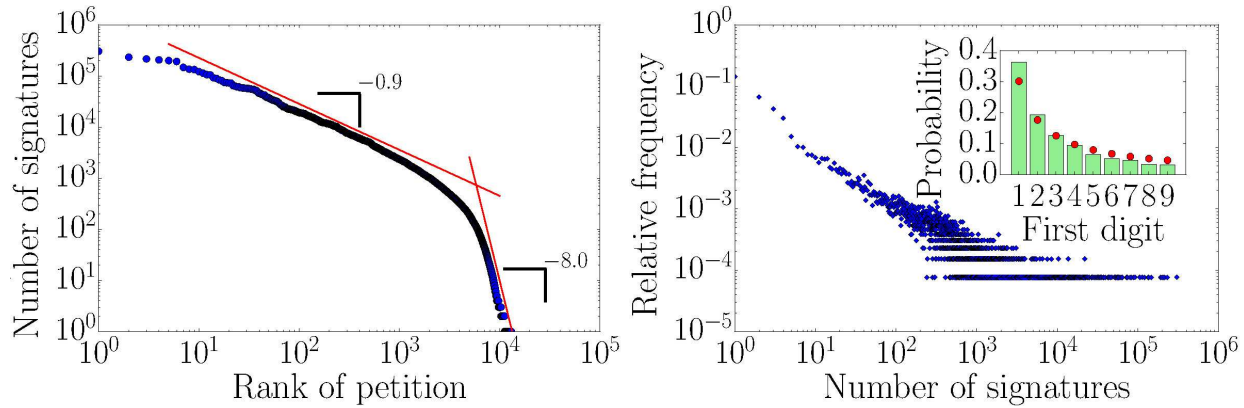}
  \caption{\textbf{Characterizing the distribution of the number of signatures a petition accrues} (left) Number of signatures as function of their rank (Zipf plot) in the \emph{openPetition} data set. The red lines are guides to the eye with slopes $-0.9$ and $-8.0$ respectively. (right) Relative frequency of petitions in the \emph{openPetition} data set with a certain number of signatures. The inset shows the distribution of the petitions' signatures first digit (green bars) and the corresponding Benford distribution data (red dots).}
 \label{fig:stats}
\end{figure}
The time series of signing events clearly exhibits bursty behavior, i.e.~there are periods of high signing activity followed by periods of low or no activity. As an example, we show the time series of the signatures per hour of the petition with the largest number of signatures in Fig.~\ref{fig:spikes} (left). The time evolution of the total number of signatures is illustrated in Fig.~\ref{fig:spikes} (right). As illustrated in the inset of Fig.~\ref{fig:spikes} (left), we also observe a circadian rhythm in the petitions' time series, which is typical for human activities. However, it is clear that bursty behavior is observed on a larger time scale. We find that the heterogeneity of inter-event time distributions holds independently of the popularity of the petition, looking at the behavior of the four different classes of numbers of signatures. However, we note that there is a systematic increase in the frequency of short inter-event times and decrease in longer inter event-times as the popularity of a petition increases.

Before discussing the actual time series analysis, we shortly analyze the petitions in our data set in terms of their number of signatures. In previous studies it has been found that only a small fraction of petitions accumulate a considerable amount of signatures \cite{yasseri13,hale13}. 
For the petitions in our data set a Zipf plot (number of signatures vs.~petition rank) is shown in Fig.~\ref{fig:stats} (left). The red lines are guides to the eye with slopes $-0.9$ and $-8.0$. The change in slope has been also observed previously for another online petition platform \cite{hale13}. However, in the latter case the change occurs at a number of signatures of around 500 which corresponds to the predefined quorum of 500 in their data set and led the authors to the conclusion that petitions are less eager
to accrue additional signatures after meeting the number required for an official government response. In our data there is no fixed quorum value since the \emph{openPetition} quorum scales with the number of inhabitants in the target region. The different slopes might be a consequence of different growth dynamics. In Fig.~\ref{fig:stats} (right) we show the probability distribution of the numbers of signatures. Around 30 \% of all petitions in our data set only have one or zero signatures. However, some petitions (0.01\%) acquired more than 100.000 signatures. %(See the \emph{Supporting Information} Tab.~S\ref{tab::vote_tab} for a description of the petitions with the highest numbers of signatures.) 
The inset in Fig.~\ref{fig:stats} (right) shows good agreement between the distribution of the numbers of signatures' first digits and the expected Benford distribution which is a typical feature of logarithmically distributed data \cite{diekmann07,diekmann10}.

\pagebreak
\subsection{Petition spike trains and local variation}
\label{sec:lv}
\begin{figure}[!htp]
\centering
\includegraphics[width=\textwidth]{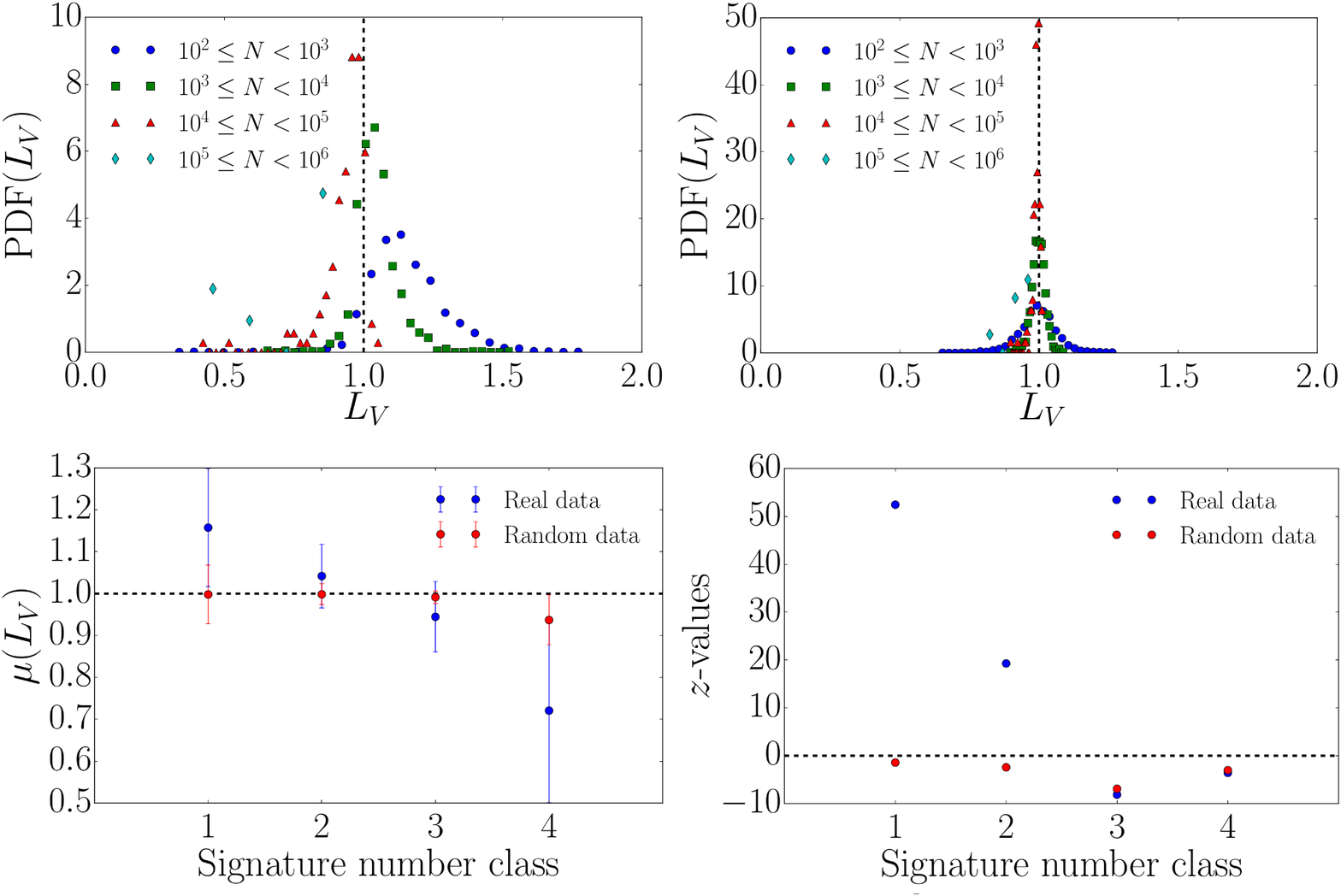}
  \caption{\textbf{Local variation analysis of petition signing spike trains for different classes of numbers of signatures.} All petitions are divided into four different classes based on the number of signatures $N$. (upper left) Distribution of the local variation for the real signing activity spike train data. (upper right) Same as the latter for randomized spike trains (null model), showing behavior that is more clearly Poissonian and the same for all classes. (lower left) The mean $\mu(L_V)$ of real and randomized spike trains for different classes of numbers of signatures. (lower right) The z-values of real and randomized data for different classes of numbers of signatures, showing that the classes with only a few signatures deviate from the Poissonian assumption according to the $L_V$ measure.}
 \label{fig:lv_panel}
\end{figure}
\begin{table}[!htp]
\caption{\textbf{F-test statistics of the local variation $L_V$.} The F-test statistics of the local variation $L_V$ in different popularity classes.}~\label{tab::ftest}
\begin{tabular}{c c c c} 
 \textit{Popularity class} & \textit{Number of petitions} & \textit{0.1-percentile} & \textit{F-value ($L_V$)} \ \\\hline\hline
$10^2\leq N < 10^3$ & 2182 & 1.04 & 2.93 \\\hline
$10^3\leq N < 10^4$ & 1213 & 1.05 & 5.39 \\\hline
$10^4\leq N < 10^5$ & 151 & 1.16 & 19.01 \\\hline
$10^5\leq N < 10^6$ & 8 & 1.76 & 48.16 \\\hline
\end{tabular}
\end{table}

To further analyze the temporal features of signing time series, we employ tools for non-stationary time series. We construct signing spike trains, i.e.~the time series of discrete signing events, for each petition. Here a spike represents signing activity at the corresponding time. We are thus not taking into account multiple signing activity---a legitimate approximation since nearly all signing activity corresponds to one signing event per time stamp, cf.~Fig.~\ref{fig:tau_pdf} (right).

As a null model for comparison, we generate a randomized sequence of the original inter-event intervals within the considered time interval, i.e.~the total number of seconds between the petition's start and end \cite{sanli15}. This procedure destroys the inter-event time correlations but preserves the distribution. Using this null model allows to study if correlations are the key factor to produce the observed local variation.
\begin{figure}[!htp]
\centering
\includegraphics[width=\textwidth]{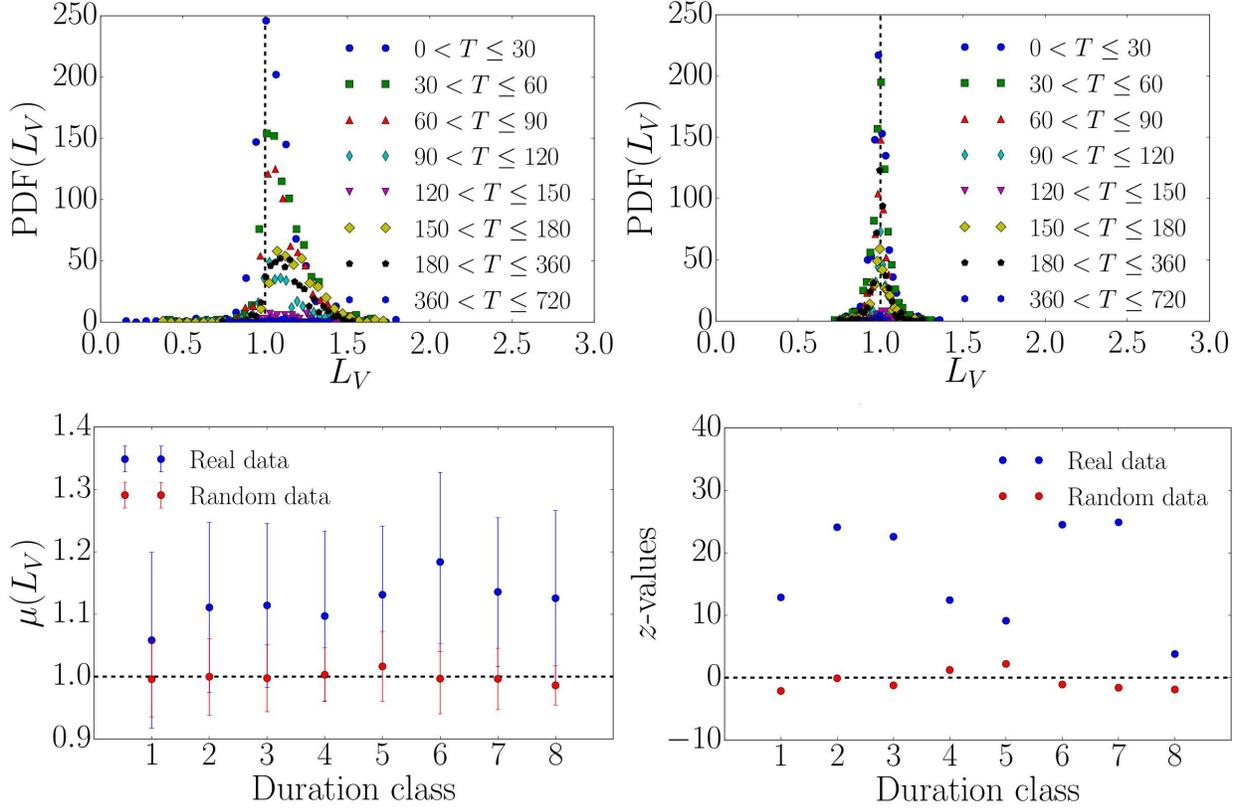}
  \caption{\textbf{Local variation analysis of petition signing spike trains for different duration classes.} All petitions are divided into eight different classes based on their duration $T$. A small class index corresponds to short durations and large one to long durations. (upper left) Distribution of the local variation for the real petition spike train data. (upper right) Same as the latter for randomized spike trains, showing behavior that is more clearly Poissonian and the same for all classes. (lower left) The mean $\mu(L_V)$ of real and randomized spike trains for different duration classes. (lower right) The $z$-values of real and randomized data for different petition duration classes.}
 \label{fig:lv_panel_duration}
\end{figure}
\begin{table}[!htp]
\caption{\textbf{Burstiness coefficient, memory coefficient and local variation of highly popular petitions.} The burstiness coefficient, memory coefficient and local variation of the eight petitions with the largest numbers of signatures.}~\label{tab::lv_b_m}
\begin{tabular}{c c c} 
 \textit{Local variation $L_V$} & \textit{Burstiness coefficient $B$} & \textit{Memory coefficient $M$} \ \\\hline\hline
0.93 & 0.96 & 0.04 \\\hline
0.92 & 0.73 & 0.36 \\\hline
0.85 & 0.99 & 0.95 \\\hline
0.82 & 0.70 & 0.32 \\\hline
0.38 & 0.83 & 0.47 \\\hline
0.40 & 0.89 & 0.36 \\\hline
0.56 & 0.73 & 0.52 \\\hline
0.90 & 0.77 & 0.39 \\\hline
\end{tabular}
\end{table}

Before analyzing the local variation $L_V$ of the petition time series, we apply a statistical F-test as suggested in Refs.~\cite{Shinomoto2009,lambiotte16} to decide whether $L_V$ consistently characterizes the time series. This means that the variance of $L_V$ across different periods in one time series should be smaller than the variance in the population of all time series. Here we subdivide each time series in 20 slices and calculate the corresponding F-values as the ratios between the variance of $L_V$ in the population of all time series and the variances across the 20 slices \cite{Shinomoto2009,lambiotte16}. For different popularity classes, we show the F-values of $L_V$ in Tab.~\ref{tab::ftest}. The F-values are significantly larger than the $0.1$-percentile values. This suggests that the variance of $L_V$ in a single time series is significantly smaller than the variance in the population and we conclude that the local variation $L_V$ properly characterizes our time series.

\begin{figure}[!htp]
\centering
\includegraphics[width=\textwidth]{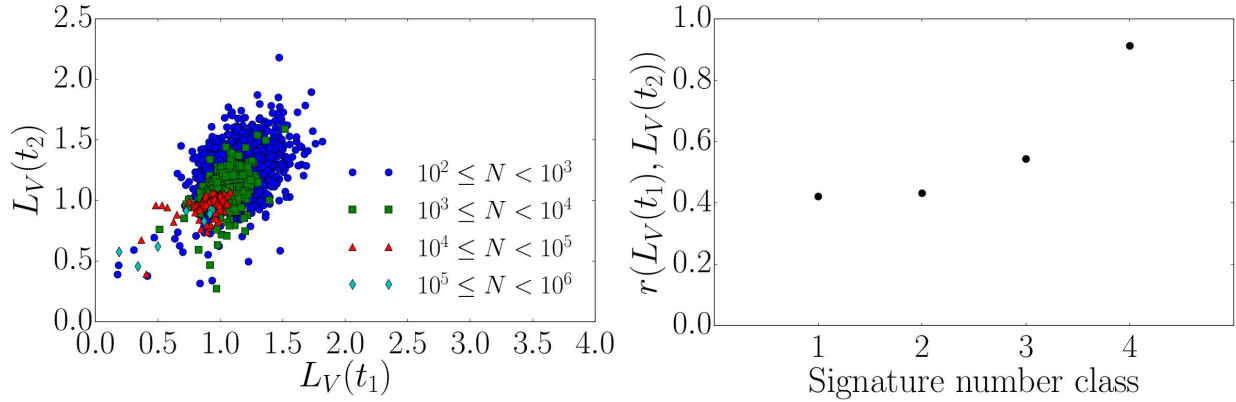}
  \caption{\textbf{Correlation of local variation for successive time intervals and the related Pearson correlation coefficient.} (left) Correlation plot for the local variation $L_V$ for two successive time intervals. (right) The corresponding Pearson correlation coefficient for different classes of numbers of signatures.}
 \label{fig:correlation}
\end{figure}
We now compute the local variation $L_V$ as defined in Eq.~\eqref{eq:lv} and see differences in the distributions of $L_V$ between the real spike train data in Fig.~\ref{fig:lv_panel} (upper left) and the randomized null model data in Fig.~\ref{fig:lv_panel} (upper right). For the eight petitions with the largest numbers of signatures, we present the corresponding $L_V$ values in Tab.~\ref{tab::lv_b_m}. Popular petitions with large numbers of signatures exhibit smaller values of $L_V$ compared to less popular ones suggesting that the former are less bursty. As described in the \emph{Materials and Methods} section, we expect $\langle L_V \rangle=1$ for a Poisson process and we find deviations for the real data as illustrated in Fig.~\ref{fig:lv_panel} (lower left). To capture the strength of the deviation
we compute the $z\text{-value}=(\mu(L_V)-\mu_0)/\sigma(L_V)/\sqrt{n}$, where $\mu_0=1$, $\mu(L_V)$ and $\sigma(L_V)$ define the mean and standard deviation of a given $L_V$ distribution respectively and $n$ is the number of data points. The $z\text{-values}$ of the randomized spike trains are almost zero indicating a Poissonian signal whereas petitions with low numbers of signatures exhibit a very large $z$-value, indicating anomalously high $L_V$, as shown in Fig.~\ref{fig:lv_panel} (lower right). This indicates that petitions with fewer signatures deviate more from the Poissonian null model. The results for $L_V$ are in agreement with a similar analysis involving $L_V$ of different popularity classes in Twitter data \cite{sanli15,sanli2015}. In Ref.~\cite{sanli15}, popular hashtags were found to exhibit less bursty spike trains compared to less popular ones and in Ref.~\cite{sanli2015} more popular user's activity is less bursty than that of sporadic users. It is worth noting that petition signing seems to more closely resemble the second case of user's activity, since the popular spike trains exhibit $L_V$ values just below 1, while in the first study $L_V \ll 1$. Interestingly, when we partition petitions by their duration, i.e.~the time period in which people have the possibility to sign a petition, we do not find that the burstiness varies according to the duration class cf.~Fig.~\ref{fig:lv_panel_duration}.

In accordance with Ref.~\cite{sanli15}, we study the persistence of $L_V$ through time by dividing each time series into two halves to calculate the local variation $L_V(t_1)$ in the first half and $L_V(t_2)$ in the second half. We find that higher numbers of signatures lead to higher correlations \cite{Shinomoto2009} between these two values of $L_V$ (see Fig.~\ref{fig:correlation}). In the past similar patterns have been observed in Twitter data \cite{sanli15}, with the critical difference that high correlation coefficients have only been found for intermediate classes. This effect might be an artifact due to the small number of samples in classes with extremely high activity.
In fact, when excluding the classes of extremely high activity the data presented in Fig.~\ref{fig:correlation} and the findings on twitter data \cite{sanli15} indicate that the distribution of $L_V$ is getting narrower with increasing activity causing an increase in correlation.
\subsection{Burstiness coefficient and the role of memory}
\begin{figure}
\centering
\includegraphics[width=\textwidth]{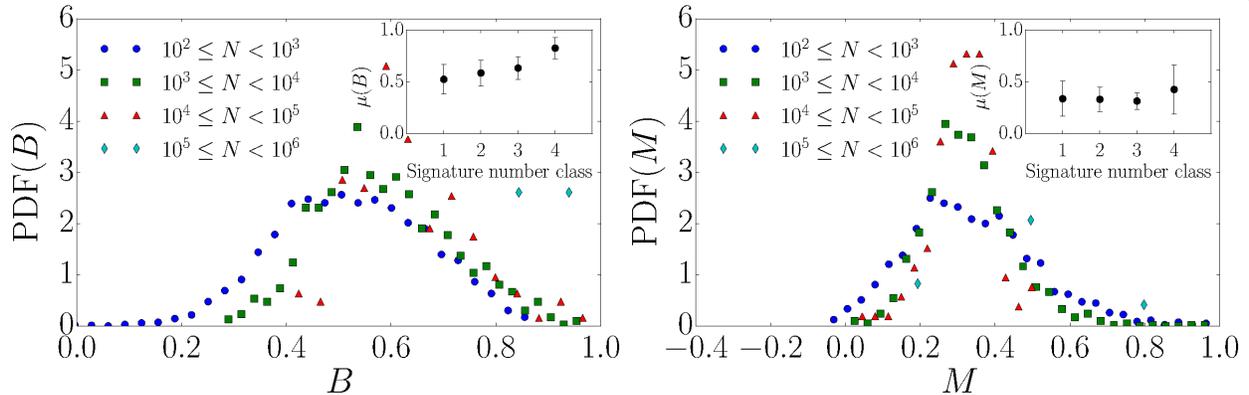}
  \caption{\textbf{Inter-event time distribution and burstiness.} (left) The PDF of the burstiness coefficient $B$ and the corresponding mean values (inset). (right) The PDF of the memory coefficient $M$ and the corresponding mean values (inset).}
 \label{fig:bpdf_mpdf}
\end{figure}
\begin{figure}
\centering
\includegraphics[width=\textwidth]{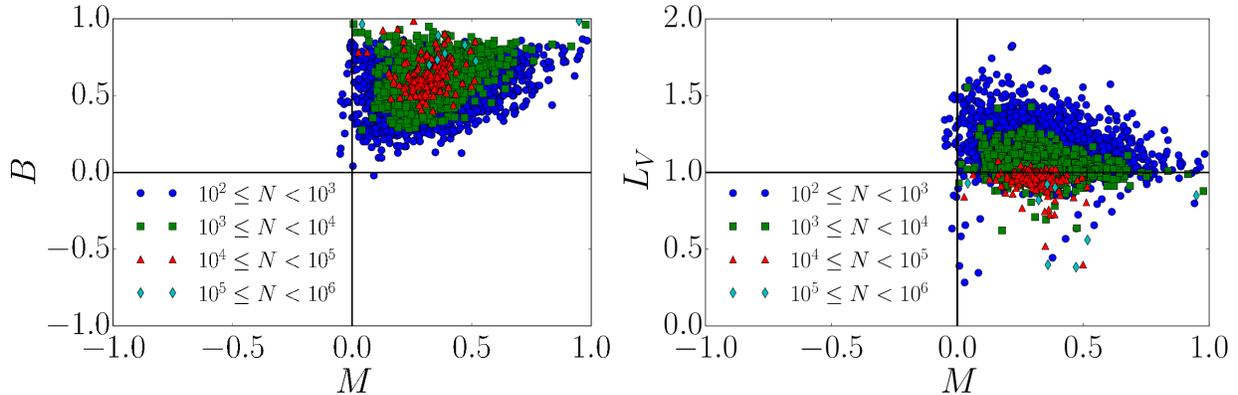}
  \caption{\textbf{Memory coefficient and burstiness.} (left) Burstiness vs.~memory coefficient. One clearly sees deviations from the human activity patterns in Ref.~\cite{goh08}, where $M$ is close to zero. (right) The local variation $L_V$ vs.~memory coefficient.}
 \label{fig:bmplot}
\end{figure}
According to the results presented in Fig.~\ref{fig:lv_panel} (upper left) that are based on the local variation $L_V$, popular petitions with a large number of signatures are less bursty compared to the ones with less signatures. However, a characterization based on the burstiness coefficient $B$ as defined in Eq.~\eqref{eq:burstiness} indicates more bursty signals for larger numbers of signatures (see inset in Fig.~\ref{fig:bpdf_mpdf} (left)).
Why do high signing petitions exhibit burstiness according to the burstiness coefficient $B$ but not according to the local variation $L_V$? This difference is due to the an important difference in the conceptual and formal definitions of burstiness according to  $L_V$ and $B$. Namely, $L_V$  takes the order of inter-event times into account, whereas $B$ does not.

The discrepancy between the effect can be partly understood measuring the memory $M$ of the inter-event time distribution, according to Eq.~\eqref{eq:memory}. For the different petition classes we illustrate their burstiness coefficient, memory coefficient, and the local variation in Fig.~\ref{fig:bmplot}. For the eight petitions with the largest numbers of signatures, we present the corresponding $B$, $M$ and $L_V$ values in Tab.~\ref{tab::lv_b_m}.
The local variation $L_V$ is based on the differences of consecutive inter-event times. However, for the computation of $B$ the order of inter-event times does not matter. Memory is a measure of the correlation between adjacent inter-event intervals. The probability density function of $M$ is illustrated separately in Fig.~\ref{fig:bpdf_mpdf} (right). We find a positive memory coefficient with an average value of $M\approx 0.3$ as also contained in Fig.~\ref{fig:bmplot} (right). This suggests that such correlations between adjacent intervals are an important source of burstiness in the petition's signing process and thus $B$ alone is not a good measure of the burstiness, as suggested in ~\cite{goh08}. Interestingly, unlike prior analysis of patterns of human activity~\cite{goh08} (e.g.~e-mail communication \cite{eckmann04}), we find a non-negligible positive memory coefficient. 
To summarize, burstiness is a consequence of strong correlations in daily human activity and thus cannot be captured by the burstiness coefficient $B$ alone. The memory coefficient $M$, as defined in Eq.~\eqref{eq:memory}, captures some of these correlations but capture differences between subsequent inter-event times. However, the local variation $L_V$ quantifies this local time order.
\section{Discussion}
In this study we focussed on the characterization of online petition time series based on data from the \emph{openPetition} online platform. The burstiness of petition signing is systematically different between those petitions that receive high signing volumes and those that do not. Our findings might therefore have implications on predicting whether a petition will be successful or not. Specifically, low signing petitions exhibit high local variation, but low correlations in local variation across successive time intervals. Conversely, high signing petitions exhibit local variation similar to that expected in a Poisson process, and the correlations between successive intervals are high. However, high signing petitions also exhibit a higher memory and burstiness coefficient than a Poisson process and than low signing petitions.
Thus, these observations can be reconciled if more popular petitions have, alongside a broad distribution of inter-event time intervals, periods where there are clusters of high frequency signing, leading to a low local variation and high memory. Our results suggest that the local variation $L_V$ precisely quantifies the local time order whereas the burstiness coefficient $B$ should not be used alone to quantify nonlinear time series.

The distribution of the local variation in different classes of signing activity are in agreement with previous findings on the popularity of microblogs on Twitter \cite{sanli15,sanli2015}. Popular hashtags were found to exhibit less bursty spike trains compared to less popular ones \cite{sanli15}and more popular user's activity is less bursty than that of sporadic users \cite{sanli2015}. Our results more closely resemble the ones of Ref.~\cite{sanli2015} since the popular spike trains exhibit $L_V$ values just below 1, while in Ref.~\cite{sanli15} $L_V \ll 1$.

Another interesting characteristic of the petitions is that they exhibit a positive memory coefficient, indicating that the duration of adjacent inter-event times is correlated. This is in contrast to the negligible memory coefficient that other studies of human activities have revealed \cite{goh08}. The unusually high memory of petition signing could be due to the influence of contagion and social influence dynamics on signing events or the effect of exogenous influence such as wide-spread media broadcasting of the petition or a related topic. One could expect to find smaller values of the memory coefficient for online petitions when countries are in political and social turmoil. This is the subject of future investigation.

\section{Acknowledgments}
\label{sec:acknowl}
We thank openPetition for providing us with the data and their continued support. We acknowledge financial support from the ETH Risk Center and ERC Advanced grants numbers FP7-319968 and FP7-3242247 of the European Research Council. This work was partially funded by the European Community's H2020 Program under the funding scheme ``FETPROACT-1-2014: Global Systems Science (GSS)'', grant agreement 641191 ``CIMPLEX: Bringing CItizens, Models and Data together in Participatory, Interactive SociaL EXploratories'' (http://www.cimplex-project.eu). We thank Konstantin Schaar for performing preliminary analyses of the signature time series during the initial phase of the project. LB thanks Jan Nagler for helpful discussions. LB also thanks Hans Herrmann for useful inputs and valuable comments.
\section{Author Contributions}
\label{sec:contr}
Conceived and designed the experiments: LB OWM DB. Performed the experiments: LB OWM. Analyzed the data: LB OWM. Wrote the paper: LB OWM DB.
\clearpage
\bibliography{refs}
\bibliographystyle{apsrev4-1}
\end{document}